\documentclass[12pt,english]{article}
                         \usepackage[T1]{fontenc}
                         \usepackage[latin1]{inputenc}
                         \usepackage{graphics}
                         \usepackage{babel}
\usepackage{epsfig}

\begin{document}

\begin{center}



{\huge Complex Trajectories and Dynamical Origin of Quantum Probability}

\vspace {1cm}

{\Large Moncy V. John}

\vspace{0.4cm}

{\large Department of Physics, St. Thomas College, Kozhencherry, Kerala 689641,\\
India.
}

\end{center}

\begin{abstract}

Complex quantum trajectories, which were first obtained from a modified de Broglie-Bohm quantum mechanics, demonstrate that Born's probability axiom in quantum mechanics originates from dynamics itself. We show that a normalisable probability density can be defined for the entire complex plane, though there may be regions where the probability is not locally conserved. Examining this for some simple examples such as the harmonic oscillator, we also find why there is no appreciable complex extended motion in the classical regime.
\end{abstract}





\section{Introduction}

Complex quantum trajectories were first obtained   and  drawn for the case of some fundamental problems,  such as the harmonic oscillator, potential step, wave packets etc., by modifying the de Broglie-Bohm (dBB) approach to quantum mechanics \cite{mvj1}.  For obtaining this representation, we shall substitute $\Psi=e^{i\hat{S}/\hbar}$ in the Schrodinger equation to obtain the quantum Hamilton-Jacobi equation (QHJE) \cite{goldstein}
and then postulate an  equation of motion similar to that of de Broglie:

\begin{equation}
m\dot{x} \equiv \frac {\partial \hat{S}}{\partial x}= \frac {\hbar
}{i} \frac {1}{\Psi}\frac {\partial \Psi}{\partial x},
\label{eq:xdot}   \end{equation}
for the particle.  The trajectories $x(t)$   are obtained by integrating this equation with respect to time and they will lie in a complex $x$-plane. It was observed that the above identification $\Psi=e^{i\hat{S}/\hbar}$  helps to utilize all the information contained in $\Psi$ while obtaining the trajectory. 

 We thus consider $x\equiv x_r+ix_i$ as a complex variable and restrict ourselves to single particles in one dimension. 
The complex eigentrajectories in the free particle, harmonic oscillator and potential step problems and complex trajectories for a wave packet solution were obtained in \cite{mvj1}. As an example, complex trajectories in the $n=1$  harmonic oscillator is shown in figure 1. These are the famous Cassinian ovals. The Jacobi lemniscate, that passes through $x_r=0$, $x_i=0$ is the special case of  these ovals.

\begin{figure}[ht] 
\centering{\resizebox {0.5 \textwidth} {0.2 \textheight }  
{\includegraphics {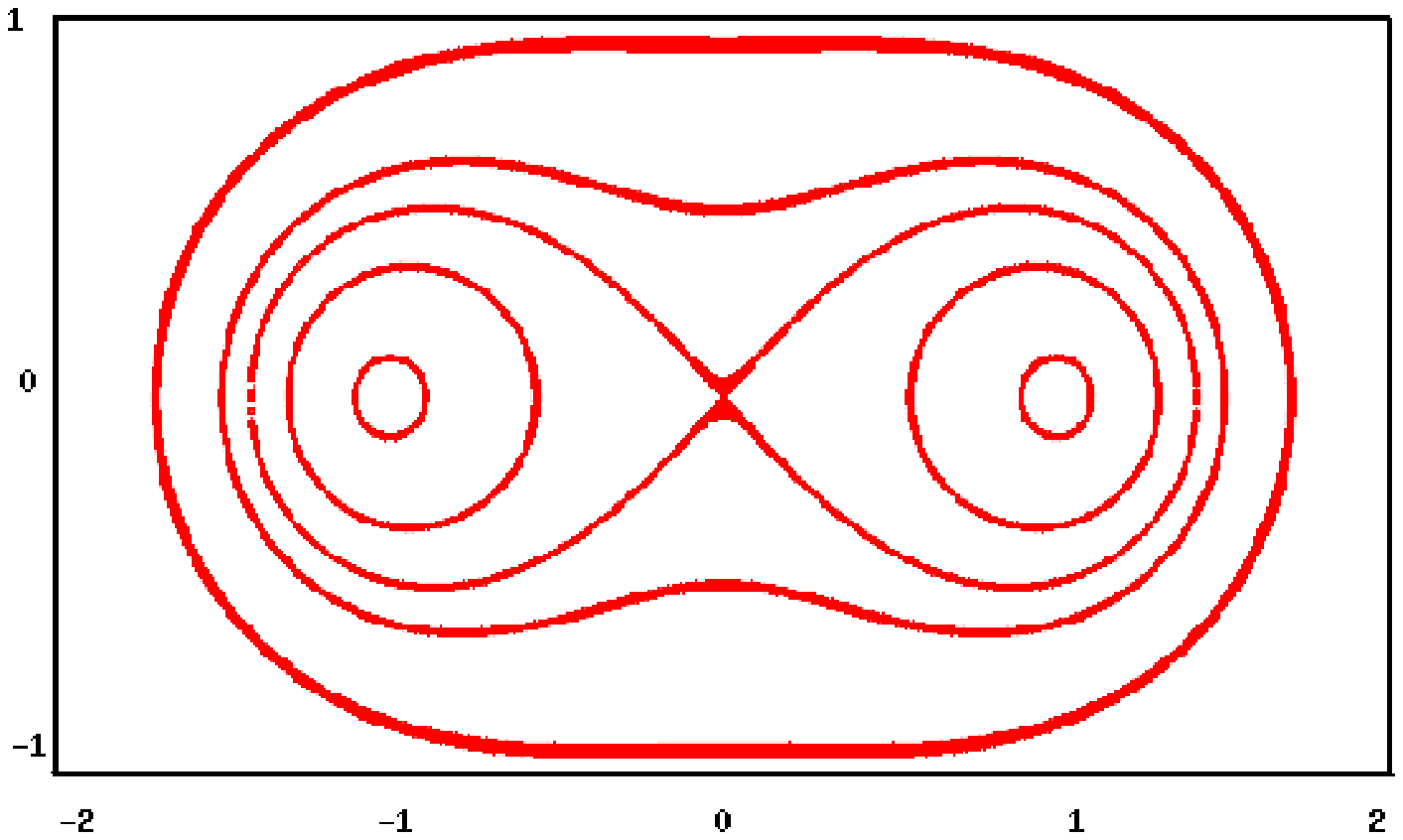}} 
\caption{ } 
 \label{fig:n1shmtraj}} 
   \end{figure}

 We may  note that even  for an eigenstate, the particle can be in any one of its infinitely many possible quantum trajectories, depending on its initial position in the complex plane. Therefore,   the expectation values of dynamical variables are to be evaluated over an ensemble of particles in all
possible trajectoriess. It was postulated that the average of a dynamical variable
$O$ can be obtained using the  measure $\Psi^{\star} \Psi $  as
$
<O> = \int_{-\infty}^{\infty} O \Psi^{\star} \Psi
 \;  dx, 
$
where the integral is to be taken along the real axis  \cite{mvj1}. Also it was noted that in this form, there is no need to make the conventional operator replacements. The above postulate is equivalent to the Born's probability axiom for observables such as position, momentum, energy, etc., and one can show that 
$<O>$ coincides with the corresponding quantum mechanical expectation
values. This makes the new scheme  equivalent to standard 
quantum mechanics when  averages of dynamical variables are computed.

However, one of the  challenges before this complex quantum trajectory representation  is to explain the  quantum probability axiom.  In a recent work which explores the connection between probability and complex quantum trajectories \cite{mvj2},  the  probability density to find the particle around some point on the real axis $x=x_{r0}$  was proposed to be

\begin{equation}
\Psi^{\star}\Psi (x_{r0},0) \equiv P(x_{r0}) ={\cal N} \exp \left({-\frac{2m}{\hbar}\int^{{x_{r0}}} \dot{x}_i dx_r}\right) , \label{eq:psistarpsi}
\end{equation}
where the integral is taken along the real line.  Since it is defined and used only along the real axis, the continuity equation for probability in the standard quantum mechanics is valid here also, without any modifications. This possibility of regaining the quantum probability distribution from the velocity field is a unique feature of the complex trajectory formulation. For instance, in the de Broglie approach, the velocity fields  for all bound eigenstates are zero everywhere  and it is not possible to obtain  a relation between velocity and probability.

At the same time, since we have the complex paths,  it would be natural to consider the probability for the particle to be  in a particular path. In addition, we may consider the probability to find the particle around different points in the same path, which can also be different. Thus it is desirable to extend the probability axiom  to the  $x_rx_i$-plane and  look for the probability of a particle to be  in an area $dx_rdx_i$ around some point ($x_r, x_i$) in the complex plane. Let this quantity be denoted as $\rho(x_r, x_i)dx_rdx_i$.

It is natural to expect that such an extended probability density agrees with Born's rule along the real line. We impose such a boundary condition for $\rho(x_r,x_i)$. Also,   it is  necessary to see whether probability conservation holds everywhere in the $x_rx_i$-plane.  It is ideal if we  have an expression for $\rho(x_r,x_i)$, which satisfies these two conditions.

Such an extended probability density was proposed in \cite{mvj2}. It was postulated that if $\rho_0$, the extended probability density at some point $(x_{r0},x_{i0})$ is given, then $\rho(x_r,x_i)$ at another point that lies on the trajectory which passes through $(x_{r0},x_{i0})$, is

\begin{equation}
\rho (x_r,x_i) = \rho_0  \exp\left[  \frac{-4}{\hbar}\int_{t_0}^{t} Im\left(\frac{1}{2}m\dot{x}^2+V(x)\right)dt^{\prime}    \right]. \label{eq:rho_def}
\end{equation}
Here, the integral is taken along the trajectory $[x_r(t^{\prime}),x_i(t^{\prime})]$. One can show that the desired continuity equation for the particle, as it moves along,
follows from it. 

While evaluating $\rho$ with the help of (\ref{eq:rho_def}) above, one needs to know $\rho_0$ at $(x_{r0},x_{i0})$ and if we choose this point as $(x_{r0},0)$, the point of crossing of the trajectory on the real line, then $\rho_0$ may  take the value $P(x_{r0})$ and may be found using  (\ref{eq:psistarpsi}).

On the other hand, if we  solve the continuity equation  for  time-independent problems with the given boundary condition, it is possible to show that the two methods give identical results.
The important property that the characteristic curves  for the continuity equation are  identical to the complex paths of particles in the present quantum trajectory representation can also be demonstrated.

\begin{figure}[ht] 
\centering{\resizebox {0.5 \textwidth} {0.2 \textheight }  
{\includegraphics {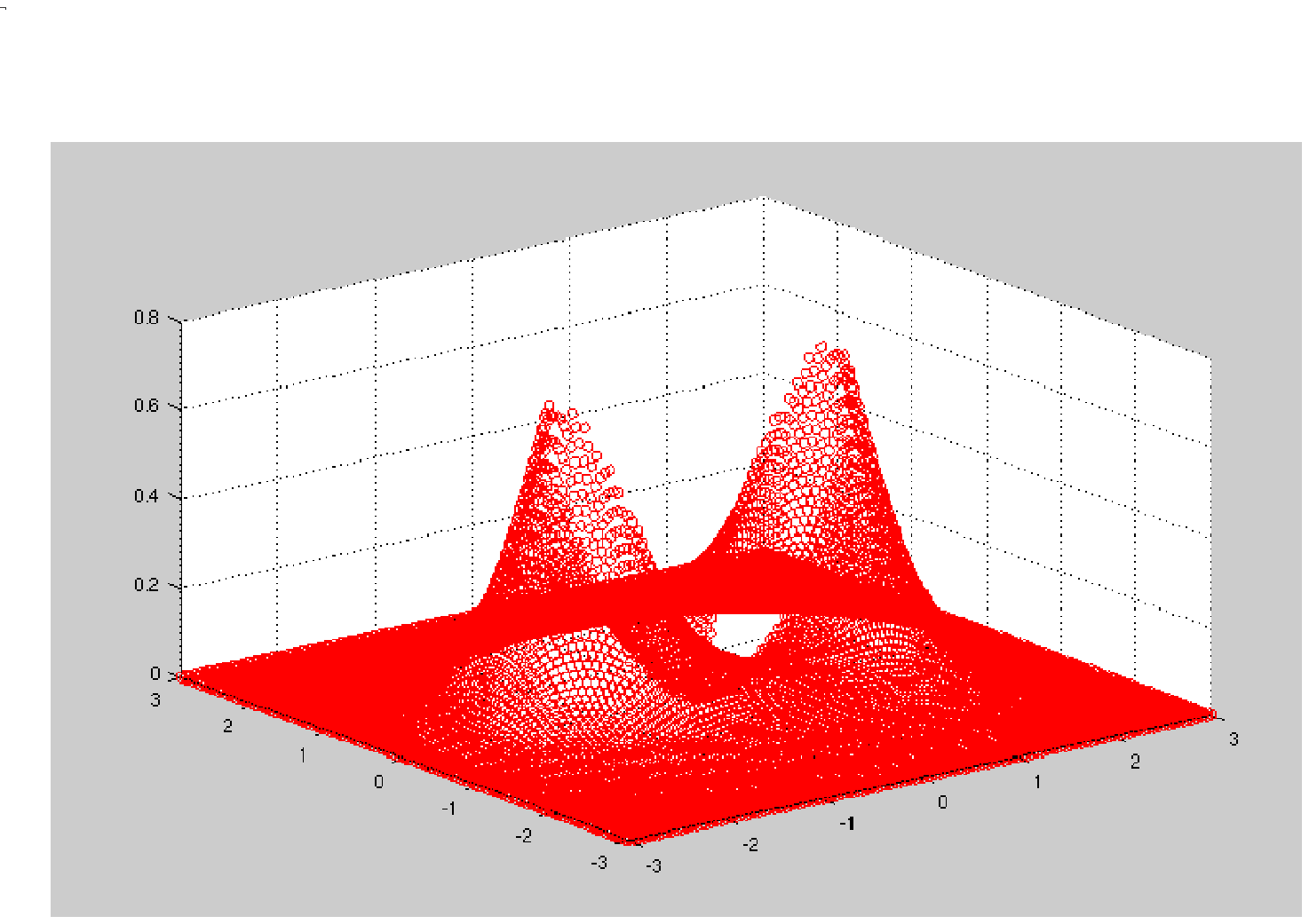}} 
\caption{ }  
\label{fig:n1shm}}  
  \end{figure}

As mentioned above,  Eq. (\ref{eq:rho_def})   gives a conserved probability along any trajectory  in the $x_rx_i$-plane. But  we have  required that the  extended probability agrees with $\Psi^{\star}\Psi$ probability along the real line.  It is seen that such an agreement is not possible for those trajectories which do not enclose the poles of $\dot{x}$. In the context of solving the conservation equation, it is easy to see that this is due to the boundary condition overdetermining the problem. In the trajectory integral approach, one can explain it as a disagreement of the values of $\rho $ at two consecutive points of crossing of the trajectory on the real axis, with that prescribed by Born's rule. Put in other words, as a particle trajectory is traversed, if the probability $\rho $ at one point of crossing  $x_{r0}$ on the real axis agrees with $P(x_{r0})$, then at the other point, say the point $x_{r1}$ where the trajectory again crosses the real line, the probability calculated according to (\ref{eq:rho_def}) will be different from that of $P(x_{r1})$. 

Given this situation,  one can ask whether it is possible to find a trajectory integral definition for $\rho$  that can agree with the Born's rule (on the real line) in this region, even if it is not conserved in the extended region. Such a definition was found \cite{mvj3} similar to that in equation (\ref{eq:rho_def}):

\begin{equation}
\rho^{\prime} (x_r,x_i) \propto P(x_{r0}) \exp\left[  \frac{-4}{\hbar}\int_{t_0}^{t} Im\left(\frac{1}{2}m\dot{x}^2\right)dt^{\prime}    \right], \label{eq:rho_alt_ls1}
\end{equation}
 The difference with the definition (\ref{eq:rho_def}) is  the absence of the potential term $V(x)$ in the integrand. This trajectory integral approach can be seen to give the same  $\Psi^{\star}(x)\Psi (x)$, the extended probability considered in \cite{wyatt3,yang3}.

\begin{figure}[ht] 
\centering{\resizebox {0.5 \textwidth} {0.2 \textheight }  
{\includegraphics {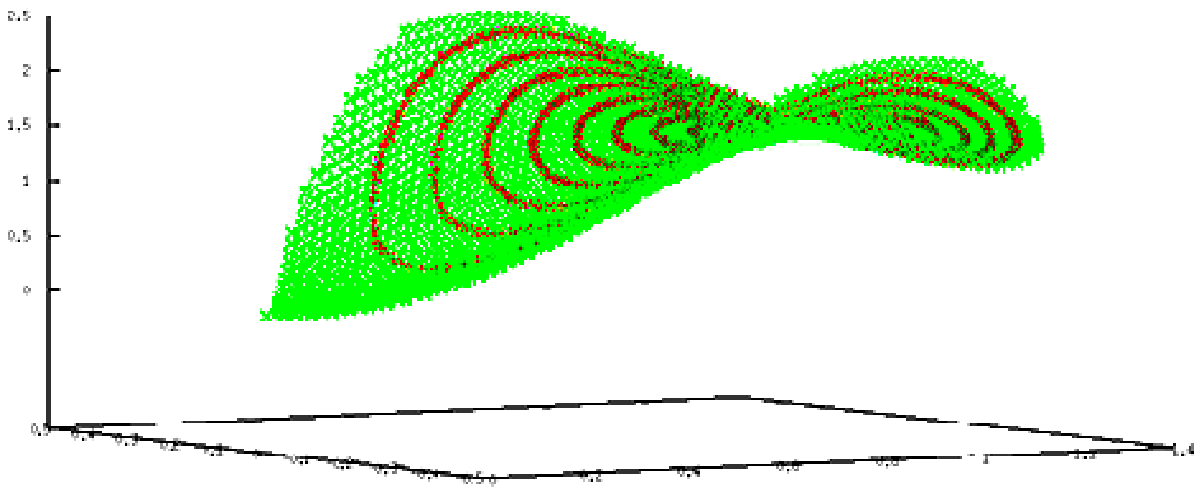}} 
\caption{  } 
 \label{fig:n1shm_ls1}}    
 \end{figure}

The extended probability density for the $n=1$ harmonic oscillator in the region inside the lemniscate (for  $x_r>0$), computed using the trajectory integral approach in (\ref{eq:rho_alt_ls1}) is shown overlapped with the `leaf-shaped' surface $\Psi^{\star}\Psi$ in this case,  in Fig. (\ref{fig:n1shm_ls1}).  We can show that the total extended probability can be normalized for the $n=1$ harmonic oscillator.

Summarising, it is seen that both the Born's probability along the real line and the extended probability in the  $x_rx_i$-plane are obtainable in terms of certain line integrals. In the  extended case, a conserved probability, which agrees with the boundary condition (Born's rule) along the real axis, is found to exist in most regions. The trajectory integral in this case is over  the imaginary part of $(1/2)m\dot{x}^2+V(x)$. For other regions (such as the region inside the lemniscate in the $n=1$ harmonic oscillator case),  an alternative definition for probability satisfies the boundary condition on the real axis, though it is not a conserved one. The integrand here is simply $(1/2)m\dot{x}^2$. Since the latter is defined only for the subnests which do not enclose the poles of $\dot{x}$, there is no difficulty in normalising the combined probability for the entire extended plane.

Another observation we make is regarding the classical limit of this complex trajectory formulation. First we note  that the probability in the region inside the lemniscate is substantial;  $43.25\%$ of the total probability lies inside this region for the $n=1$ harmonic oscillator. The maximum value of $x_i$ for the lemniscate  (a measure of its width) in this case  is $x_i^{max}=  X_i^{max}/\alpha =0.4858/\alpha$. This explains how  classical particles are confined close to the real line. For instance, consider a classical harmonic oscillator of mass $m\sim 1$ kg and frequency $\sim 1$ Hz  in the $n=1$ state. The maximum value of $x_i$ for its lemniscate is $x_i^{max}\approx 10^{-17}/\sqrt{m\omega_0}$ in units of metres. Thus $43.25\%$ of the total probability in this case lies within the lemniscate of  width $\sim 10^{-17}$ m. It may also be noted that since the extended probability outside the lemniscate decreases fast for larger ovals,  most of the probability outside  also lies close to the real line. 

For the higher energy eigenstates of the harmonic oscillator,  the lemniscates are seen to be  narrower than that of the low energy ones.  It can be assured that for a classical oscillator of any energy and having mass 1 kg., the width of the region, where the large part of probability lies, is less than or of the order of $10^{-17}/\sqrt{\omega_0}$ m. This indicates that the probable imaginary part of position for classical particles are of  extremely small size. However, we may also see that  an electron executing harmonic oscillation with frequency $\omega_0$  has to accommodate relatively large values for $x_i$, approximately equal to $ 0.01/\sqrt{\omega_0}$ m. In summary, the probability axiom in the modified de Broglie quantum mechanics helps to distinguish  the classical limit of quantum harmonic oscillator as one in which the oscillator is probable to be found only  very close to the real axis. This  result is very important for complex quantum trajectories, for it explains  why the complex extension is not observable even indirectly in the classical limit.  We anticipate that this property is generally  true.

\bibliographystyle{ccp6}

\end{document}